\documentclass[12pt]{article}
\usepackage{amsmath,amssymb}
\usepackage{graphicx,psfrag,epsf}
\usepackage{enumerate}
\usepackage{natbib}
\usepackage[hidelinks]{hyperref}
\usepackage{todonotes} 
\usepackage{epstopdf}

\newcommand{\blind}{1}

\begin{document}

	\def\spacingset#1{\renewcommand{\baselinestretch}%
		{#1}\small\normalsize} \spacingset{1}

	\if1\blind
	{
		\title{\bf Informed Bayesian $T$-Tests}
		\author{Quentin F.~Gronau$^1$\thanks{
				Correspondence concerning this article should be addressed to: Quentin F. Gronau, University of Amsterdam, Nieuwe Achtergracht 129 B, 1018 WT Amsterdam, The Netherlands. E-mail may be sent to Quentin.F.Gronau@gmail.com. This research was supported by a Netherlands Organisation for Scientific Research (NWO) grant to QFG (406.16.528) and by a Vici grant from the NWO to EJW (016.Vici.170.083), which also funded AL. AL was in part funded by the research program NWO TOP ``Safe Bayesian Learning'' with project number 617.001.651. Centrum Wiskunde \& Informatica (CWI) is the national research institute for mathematics and computer science in the Netherlands. We thank the editor, the associate editor, and two reviewers for their constructive suggestions for improvement. The authors are also grateful for the suggestions by Richard Morey on an earlier draft. The authors would also like to express their gratitude to Suzanne Oosterwijk for taking part in the prior elicitation effort. The authors would furthermore like to thank Mich{\`e}le Nuijten for her help with obtaining the frequency of tests reported in \citet{nuijten2016prevalence}. \texttt{R} code and the online appendix can be found on the Open Science Framework: \url{https://osf.io/37vch/}.}\hspace{.2cm} \\
			Alexander Ly$^{1,2}$\\
			and \\
			Eric-Jan Wagenmakers$^1$\vspace{.3cm}\\
			$^1$University of Amsterdam \\
			Department of Psychological Methods\\
			The Netherlands\\[0.2cm]
			$^2$Centrum Wiskunde \& Informatica\\
			The Netherlands\\}
		\date{}
		\maketitle
	} \fi
	
	\if0\blind
	{
		\bigskip
		\bigskip
		\bigskip
		\begin{center}
			{\LARGE\bf Informed Bayesian $T$-Tests}    
		\end{center}
		\medskip
	} \fi
	
	\bigskip
	\begin{abstract}
		Across the empirical sciences, few statistical procedures rival the popularity of the frequentist \( t \)-test. In contrast, the Bayesian versions of the \( t \)-test have languished in obscurity. In recent years, however, the theoretical and practical advantages of the Bayesian \( t \)-test have become increasingly apparent and various Bayesian $t$-tests have been proposed, both objective ones (based on general desiderata) and subjective ones (based on expert knowledge). Here we propose a flexible $t$-prior for standardized effect size that allows computation of the Bayes factor by evaluating a single numerical integral. This specification contains previous objective and subjective $t$-test Bayes factors as special cases.
		Furthermore, we propose two measures for informed prior distributions that quantify the departure from the objective Bayes factor desiderata of predictive matching and information consistency. We illustrate the use of informed prior distributions based on an expert prior elicitation effort.
	\end{abstract}
	
	\noindent%
	
	\newpage
	\spacingset{1.45} 
	
	\section{INTRODUCTION}
	\label{informedSecIntro}
	The $t$-test is designed to assess whether or not two means differ. The question is fundamental, and consequently the $t$-test has grown to be an inferential workhorse of the empirical sciences. The popularity of the $t$-test is underscored by considering the $p$-values published in eight major psychology journals from 1985 until 2013 \citep{nuijten2016prevalence}; out of a total of 258,105 $p$-values, 26\% tested the significance of a $t$ statistic. For comparison, 4\% of those $p$-values tested an $r$ statistic, 4\% a $z$ statistic, 9\% a $\chi^{2}$ statistic, and 57\% an $F$ statistic. Similarly, \citet{wetzels2011statistical} found 855 $t$-tests reported in 252 psychology articles, for an average of about  3.4 $t$-tests per article. 
	
	The two-sample $t$-test typically assumes that the data are normally distributed with common standard deviation, that is, $Y_{1i} \sim \mathcal{N} ( \mu + \tfrac{ \sigma \delta}{2}, \sigma^{2})$ and $Y_{2j} \sim \mathcal{N} ( \mu - \tfrac{ \sigma \delta}{2}, \sigma^{2})$ for $i=1, \ldots, n_{1}$ and $j=1, \ldots, n_{2}$. The parameter $\mu$ is interpreted as a grand mean, $\sigma$ as the common standard deviation, and $\delta$ as the (standardized) effect size. A typical application involves a treatment group and a control group and the task is to infer whether or not the treatment has an effect. The null hypothesis of the treatment not being effective corresponds to $\mathcal{H}_{0}: \delta = 0$ and implies that the population means of the two groups are the same, while the two-sided alternative $\mathcal{H}_1$ allows the effect size to vary freely, and implies that the population means of the two groups differ.
	
	This article concerns the Bayesian $t$-test originally developed by \citet{jeffreys1948theory2} in the one-sample setting, and recently extended to the two-sample set-up by \citet{gonen2005bayesian} and, subsequently, \citet{rouder2009bayesian}. 
	In his work on hypothesis testing, Jeffreys focused on the \emph{Bayes factor} \citep{etz2017haldane, kass1995bayes, ly2016harold, ly2016evaluation, RobertEtAl2009}, the predictive updating factor that quantifies the change in relative beliefs about the hypotheses $\mathcal{H}_1$ and $\mathcal{H}_0$ based on observed data $d$ \citep[][p. 387]{wrinch1921certain}:
	\begin{align}
		\label{informedEqPostmodelratio}
		\underbrace{\frac{P(\mathcal{H}_{1} \, | \, d)}{P(\mathcal{H}_{0} \, | \,d)}}_{\text{Posterior odds}} = %
		\underbrace{\frac{p(d \, | \, \mathcal{H}_{1})}{p(d \, | \, \mathcal{H}_{0})}}_{\text{BF}_{10}(d)} %
		\underbrace{\frac{P(\mathcal{H}_{1})}{P(\mathcal{H}_{0})}}_{\text{Prior odds}}.
	\end{align}
 	The Bayes factor is given by the ratio of the marginal likelihoods of $\mathcal{H}_1$ and $\mathcal{H}_0$ that are obtained by integrating out the model parameters with respect to the parameters' prior distribution.		
	For the two-sample $t$-test, the null model $\mathcal{H}_{0}$ specifies two free parameters $\zeta = ( \mu, \sigma)$, while the alternative has three, namely, $(\zeta, \delta) = (\mu, \sigma, \delta)$. Once the priors $\pi_{0}(\zeta)$ and $\pi_{1}(\zeta, \delta)$ are specified, the parameters of each model can be integrated out as follows
	\begin{equation}
		\label{informedEqBF}
		\text{BF}_{10}(d) = \frac{\int_{\Delta} \int_{Z} f(d \, | \, \delta, \zeta, \mathcal{H}_{1}) \, \pi_{1}(\delta, \zeta) \, \textnormal{d} \zeta \, \textnormal{d} \delta }{\int_{Z} f(d \, | \,  \zeta, \mathcal{H}_{0}) \, \pi_{0}(\zeta) \, \textnormal{d}  \zeta}.
	\end{equation}
	Eq.~\ref{informedEqBF} shows that the Bayes factor can be regarded as the ratio of two weighted averages where the weights correspond to the prior distribution for the parameters. Consequently, the choice of the prior distributions is crucial for the development of a Bayes factor hypothesis test. \citet{jeffreys1961theory} elaborated on various procedures to select priors for a Bayes factor and the construction of his one-sample \( t \)-test became the norm in objective Bayesian analysis (e.g., \citealp{bayarri2012criteria,BergerPericchi2001,liang2008mixtures}). Jeffreys's Bayes factor for the two-sample $t$-test, however, was needlessly complicated and it was \citet{gonen2005bayesian} who provided the desired simplification.

   The innovation of \citet{gonen2005bayesian} was to reparameterize the means of the two groups, \( \mu_{1} \) and \( \mu_{2} \), in terms of a grand mean and the effect size, as was introduced at the start of this section. Following Jeffreys, the second idea was to use a right Haar prior $\pi_{0}(\mu, \sigma) \propto \sigma^{-1}$ on the nuisance parameters, the parameters common to both the null and the alternative model (\citealp{bayarri2012criteria}, \citealp{berger1998bayes}, \citealp{severini2002exact}). Using this prior choice, the marginal likelihood of the null model --the denominator of the Bayes factor $\text{BF}_{10}(d)$-- is proportional to the density of a standard $t$-distribution evaluated at the observed $t$-value. The third idea was to decompose the prior under the alternative hypothesis into a product of the prior used under the null hypothesis, and a test-relevant prior on the (standardized) effect size, that is, $\pi_{1} ( \mu, \sigma, \delta) = \pi_{0}(\mu, \sigma) \pi(\delta)$. Finally, \citet{gonen2005bayesian} showed that a normal prior $\delta \sim \mathcal{N} ( \mu_{\delta}, g)$ on the effect size yields a Bayes factor for the two-sample $t$-test that is easily calculated:
   	\begin{equation}
   	\label{eqGonen}
   	\text{BF}_{10}(d;  \mu_{\delta}, g) = \frac{ \tfrac{1}{\sqrt{1+ n_{\delta} g}} T_{\nu} ( \tfrac{t}{\sqrt{1+ n_{\delta} g}} \, ; \, \sqrt{\frac{n_{\delta}}{1+ n_{\delta} g}} \mu_{\delta})}{ T_{\nu} ( t)} ,
   	\end{equation}
   	where $\tfrac{1}{b} T_{\nu}( \tfrac{t}{b} \, ; \, a )$ denotes the density of a $t$-distribution with $\nu$ degrees of freedom, non-centrality parameter $a$ and scale $b$, $T_{\nu} (t) = T_{\nu} (t \, ; \, 0)$ denotes the density of a standard $t$-distribution, and $d$ refers to the data consisting of degrees of freedom $\nu = n_{1} + n_{2} -2$, the observed $t$-value $t = \sqrt{n}_{\delta} ( \bar{y}_{1} - \bar{y}_{2}) / s_{p}$, where $n_{\delta} = ( 1/n_{1} + 1/n_{2})^{-1}$ is the effective sample size, and $\nu s_{p}^{2} = (n_{1} - 1) s^{2}_{1} + (n_{2} - 1) s_{2}^{2}$ the pooled sums of squares.\footnote{In fact, the Bayes factors for the two-sample $t$-test discussed here also cover the one-sample case, by (1) replacing the effective sample size by the sample size $n$; (2) replacing the degrees of freedom $\nu$ by $n -1$; and (3) replacing the two-sample $t$-value by its one sample equivalent $t = \sqrt{n} \bar{y} / s_{y}$, where $\nu s_{y}^2 = \sum_{i=1}^{n} (y_{i} - \bar{y})^{2}$.}
   	This means that practitioners who can calculate a classical $t$-test can also easily conduct a Bayesian two-sample $t$-test: they only need to choose the hyperparameter $\mu_{\delta}$ corresponding to the effect size prior mean and the hyperparameter $g$ corresponding to the prior variance. For brevity, we refer to the latter choice $\delta \sim \mathcal{N} ( \mu_{\delta}, g)$ as a $g$-prior on $\delta$, since it resembles the priors \citet{zellner1986assessing} proposed in the regression framework.\footnote{When $\mu_{\delta}=0$, the normal $g$-prior on $\delta$ translates to Zellner's $g$-prior on the mean difference $(\mu_{1} - \mu_{2}) \sim \mathcal{N} (0, g \sigma^{2})$.}
   	
   	Later Bayes factors for the two-sample $t$-test proposed by \citet{rouder2009bayesian} and \citet{wang2016simple} retained the first three ideas: the parameterization in terms of the grand mean and effect size, the use of the right Haar prior on the nuisance parameters $\pi_{0} (\mu, \sigma) \propto \sigma^{-1}$, and the decomposition $\pi_{1} ( \mu, \sigma, \delta) = \pi_{0}(\mu, \sigma) \pi(\delta)$, but they differ in the choice of the test relevant prior $\pi(\delta)$. \citet{wang2016simple} noted that the Bayes factors of \citet{gonen2005bayesian} are \emph{information inconsistent}, which implies that the Bayes factor in favor of the alternative does not go to infinity when the observed $t$-value increases indefinitely.
   	To make the Bayes factor information consistent, \citet{wang2016simple} instead proposed to assign $g$ a Pearson type VI/beta prime hyper-prior distribution \citep[see also][for this proposal in the regression context]{maruyama2011fully}. Inspired by the developments of \citet{liang2008mixtures} in the regression framework, \citet{rouder2009bayesian} proposed to replace the normal prior on $\delta$ by a Cauchy prior $\pi(\delta) =\text{Cauchy}(\delta \, ; \, 0, \gamma)$, a choice that resembles that of \citet{jeffreys1948theory2} proposition for the one-sample $t$-test with prior scale $\gamma=1$. In their response to  \citet{wang2016simple}, \citet{gonen2017comparing} stressed the relevance of a subjective prior specification and noted that the Bayes factors proposed by \citet{rouder2009bayesian} and \citet{wang2016simple} are not flexible enough to incorporate available expert knowledge, since these objective Bayes factors are based on priors that are centered at zero. Here --without taking sides in the discussion between objective and subjective inference-- we present a generalized form of the Bayes factor developed by  \citet{rouder2009bayesian} that allows the prior specification to be informed by substantive domain knowledge.
   	
    The remainder of this article is organized as follows: Section 2 presents the proposed Bayes factor and two measures for quantifying the departure from Jeffreys's desiderata of predictive matching and information consistency. Section 3 demonstrates, using a concrete example, how the proposed Bayes factor can be used in practice to incorporate expert knowledge based on a prior elicitation effort. The article ends with concluding comments.
   
   \section{THEORY}
	We use the framework of \citet{gonen2005bayesian} and extend the priors proposed by \citet{rouder2009bayesian} to allow for more informed Bayesian $t$-tests. We exploit the fact that, with $\pi_{0} (\mu, \sigma) \propto \sigma^{-1}$, the Bayes factor can be written as\footnote{A derivation is provided in the online appendix (Theorem~A.1, Theorem~A.2, and the associated corollaries).}	
	\begin{equation}
		\label{eq:BF10}
		\text{BF}_{10} (d) =   \frac{ \int T_{\nu} ( t \, | \, \sqrt{n}_{\delta} \delta )  \pi(\delta) \textnormal{d} \delta}{T_{\nu}(t)},
	\end{equation}
	where $T_{\nu} (t \, | \, a)$ denotes the density of a $t$-distribution with $\nu$ degrees of freedom and non-centrality parameter $a$.
	The numerator can be easily evaluated using numerical integration.
	Consequently, Eq.~\ref{eq:BF10} shows that researchers can easily obtain a Bayes factor based on any proper prior for the standardized effect size $\delta$ by inserting the prior density of interest for $\pi(\delta)$.
		
	We propose the use of a flexible $t$-prior for $\delta$, that is, $\pi(\delta)=\tfrac{1}{\gamma} T_{\kappa} ( \tfrac{\delta - \mu_{\delta}}{\gamma})$, allowing practitioners to incorporate expert knowledge about standardized effect size by specifying a location hyperparameter $\mu_{\delta}$,  a scale hyperparameter $\gamma$, and a degrees of freedom \sloppy hyperparameter $\kappa$. The resulting Bayes factor is given by:
	\begin{equation}
	\label{eqBfInformed}
	\text{BF}_{10}(d; \mu_{\delta}, \gamma, \kappa) = \frac{\int T_{\nu}( t \, | \, \sqrt{n_{\delta}} \delta) \tfrac{1}{\gamma} T_{\kappa}( \tfrac{\delta - \mu_{\delta}}{\gamma}) \textnormal{d} \delta}{T_{\nu}(t)} , 
	\end{equation}
	where the integral in the numerator can be easily calculated using free software packages such as \texttt{R} \citep{R}. 
	We believe that the proposed Bayes factor based on a $t$-prior for effect size has a number of advantages.
	First, similar to the Bayes factor proposed by \citet{gonen2005bayesian} --which is a special case obtained by taking $\gamma = \sqrt{g}$ and $\kappa \rightarrow \infty$-- it allows researchers, if desired, to incorporate existing expert knowledge about effect size into the prior specification furthering cumulative scientific learning.
	Second, this class of priors  contains the Cauchy prior of \citet{rouder2009bayesian} as a special case (obtained by setting $\kappa=1$, $\mu_{\delta}=0$). Therefore, using the same expression, researchers can incorporate expert prior knowledge or they can use an objective default prior.
	Third, this set-up allows researchers to quantify the departure from Jeffreys's predictive matching and information consistency desiderata based on departure measures proposed below.
	This enables a more formal assessment of differences between objective and subjective prior choices and may benefit the dialog between objective and subjective Bayesians (see, e.g., \citealp{wang2016simple}, and \citealp{gonen2017comparing}).
	
	\subsection{Two measures for the departure from Jeffreys's desiderata}
	
	\subsubsection{Predictive matching}
	Jeffreys considered two desiderata for prior choice.
	The first desideratum, \emph{predictive matching}, states that the Bayes factor should be perfectly indifferent (i.e., $\text{BF}_{10}(d) = 1$) in case the data are completely uninformative.
	Recall that the alternative model has three free parameters; it is therefore natural to require at least three observations before conclusions can be drawn. Consequently, Jeffreys required a Bayes factor of 1 for any data set of size smaller or equal to 2, thus, for $\nu=0$. As apparent from Eq.~\ref{informedEqPostmodelratio}, this requirement guarantees the posterior model odds to be the same as the prior model odds for completely uninformative data sets. For instance, the data set $d_{\nu < \min}$ consisting of only one observation in each group $n_{1}=n_{2}=1$ automatically has zero sums of squares, that is, $\nu s_{p}^{2}=0$. If $\bar{y}_{1} \neq \bar{y}_{2}$ the associated $t$-value would then be unbounded.
	Let $f(d \, | \, \delta)$ denote the \emph{reduced} likelihood (i.e., the likelihood with the nuisance parameters integrated out): $f(d \, | \, \delta) = \int \int f( d\, | \, \mu, \sigma, \delta) \sigma^{-1} \textnormal{d} \mu \textnormal{d} \sigma$. Using a lemma distilled from the Bateman project \citep{bateman1954integral1, bateman1953higher2, ly2018analytic}, straightforward but tedious computations show that $f(d \, | \, \delta)$ is proportional to the density of a \( t \)-distribution with $\nu$ degrees of freedom and non-centrality parameter $\sqrt{n_{\delta}} \delta$ (see Theorem~A.2 in the online appendix for details). To convey that nothing is learned from the data set $d_{\nu < \min}$, Jeffreys chose $\pi(\delta)$ such that 
	\begin{align}
		p(d_{\nu < \min} \, | \, \mathcal{H}_{0}) = p(d_{\nu < \min} \, | \, \mathcal{H}_{1}) = \int f(d_{\nu < \min} \, | \, \delta) \pi(\delta) \textnormal{d} \delta.
	\end{align}
	As $\nu s_{p}^{2}=0$, $n_{\delta} = 1/2$, and $\bar{y}_{1} \neq \bar{y}_{2}$, we obtain
	\begin{align}
		\label{eqPredictiveMatchingRequire}
		(2  |\bar{y}_{1} - \bar{y}_{2}|)^{-1} = \int (2 | \bar{y}_{1} - \bar{y}_{2} |)^{-1} \big [ 1 + \textnormal{sign} ( \bar{y}_{1} - \bar{y}_{2}) \textnormal{Erf} ( \tfrac{\delta}{2}) \big ] \pi (\delta) \textnormal{d} \delta ,
	\end{align}
	where $\textnormal{sign}( z)$ is one when $z$ is positive, minus one when $z$ is negative, and zero otherwise (see Corollary~A.1.3 and Corollary~A.2.1 in the online appendix). $\textnormal{Erf}(z) = \tfrac{2}{\sqrt{\pi} } \int_{0}^{z} e^{- u^{2}} \textnormal{d} u$ is the error function, an odd function of $z$. Note that the requirement Eq.~\ref{eqPredictiveMatchingRequire} is fulfilled if a proper symmetric prior is used for $\delta$. Based on Eq.~\ref{eqPredictiveMatchingRequire} we define the (two-sided) departure of any proper prior with respect to Jeffreys's predictive matching criterion as
	\begin{align}
		D( \pi, \text{Pred} \, | \, d_{\nu < \min} ) = \int \textnormal{sign}( \bar{y}_{1} - \bar{y}_{2} ) \textnormal{Erf} ( \tfrac{\delta}{2} ) \pi (\delta) \textnormal{d} \delta,
	\end{align}
	and note that $\text{BF}_{10}(d_{\nu < \min}) = 1 + D( \pi, \text{Pred} \, | \, d_{\nu < \min} )$. For instance, a $t$-prior located at $\mu_{\delta} = 0.350$, with scale $\gamma = 0.103$ and $\kappa=3$ degrees of freedom, as used later on in the example, has a departure of the predictive matching criterion of 0.0198 when $\bar{y}_{1} > \bar{y}_{2}$. In other words, for completely uninformative data sets with $\bar{y}_{1} < \bar{y}_{2}$ the Bayes factor will be $\text{BF}_{10}(d_{\nu < \min}) \approx 0.98$, while if $\bar{y}_{1} > \bar{y}_{2}$ the Bayes factor would be $\textnormal{BF}_{10}(d_{\nu < \min}) \approx 1.02$, instead.
	
	\subsubsection{Information consistency}
	The second desideratum, \emph{information consistency}, states that the Bayes factor should provide infinite support for the alternative in case the data are overwhelmingly informative \citep{bayarri2012criteria, jeffreys1942significance}.
	An overwhelmingly informative data set for the two-sample $t$-test is denoted by $d_{\textnormal{info}, \nu}$ with $\nu \geq 1$, effective sample size $n_{\delta} > 1/2$,%
	\footnote{This condition implies that there is at least one observation per group.} %
	a (pooled) sums of squares $\nu s_{p}^{2} =0$, and an observed mean difference $\bar{y}_{1} - \bar{y}_{2} \neq 0$, thus, an unbounded $t$-value. For such an overwhelmingly informative data set $d_{\textnormal{info}, \nu}$ to provide infinite support for the alternative, Jeffreys required that $p(d_{\textnormal{info}, \nu} \, | \, \mathcal{H}_{0})$ is bounded and that $\pi(\delta)$ is chosen such that $\int f(d_{\textnormal{info}, \nu} \, | \, \delta) \pi(\delta) \textnormal{d} \delta$ diverges. With $\nu s_{p}^{2}=0$ and $\bar{y}_{1} \neq \bar{y}_{2}$ the marginal likelihood of the null model becomes
	\begin{equation}
		p(d_{\textnormal{info}, \nu} \, | \, \mathcal{H}_{0}) = \frac{ \Gamma ( \tfrac{ \nu + 1}{2}) }{ 2 \pi^{\tfrac{ \nu+1}{2}} \sqrt{ \nu +2}} \big ( n_{\delta} ( \bar{y}_{1} - \bar{y}_{2})^{2} \big )^{ - \tfrac{ \nu +1}{2}} ,
	\end{equation}
	which is indeed bounded (see Corollary~A.1.3 in the online appendix). In Corollary~A.2.2 of the online appendix it is shown that for $\delta$ large, the reduced likelihood $f(d_{\textnormal{info}, \nu} \, | \, \delta)$ with $\nu s_{p}^{2}=0$ behaves like a polynomial with leading order $\nu$, that is, 
	\begin{equation}
		f(d_{\textnormal{info}, \nu} \, | \, \delta) \sim \delta^{\nu}.
	\end{equation}
	To guarantee for degrees of freedom $\nu$ that $\int f(d_{\textnormal{info}, \nu} \, | \, \delta) \pi(\delta) \textnormal{d} \delta$ diverges, it suffices to take a prior that does not have the $\nu$th moment. As information consistency should hold for all $\nu \geq 1$, this implies that $\pi(\delta)$ should be chosen such that it does not have a first moment. Based on the condition that the marginal likelihood should already diverge for $\nu =1$, we define the departure of Jeffreys's information consistency criterion as 
	\begin{equation}
		D( \pi, \text{InfoConsist}) = \text{arg} \min \left \{ \nu \in \mathbb{N} \, : \, \int f(d_{\textnormal{info}, \nu} \, | \, \delta) \pi(\delta) \textnormal{d} \delta \not \in \mathbb{R} \right \} - 1.
	\end{equation}
	If $\pi(\delta)$ is taken to be a $t$-prior with $\kappa$ degrees of freedom the departure from Jeffreys's information consistency criterion is $\kappa -1$, since a $t$-distribution has $\kappa - 1$ moments. For instance, a $t$-prior with $\kappa=3$ degrees of freedom has only two moments and, therefore, misses the information consistency by two samples. This means that the Bayes factor only goes to infinity for overwhelmingly informative data when $\nu \geq 3$.
	Therefore, an informed $t$-prior with degrees of freedom larger than one requires more observations to be ``convinced'' by the data than does an objective prior with degrees of freedom equal to 1.
	
	\subsubsection{Practical value of the proposed departure measures}
    The departure measures introduced above can be used to issue recommendations for researchers who would like to incorporate expert knowledge into the prior specification, but would also like to retain Jeffreys's desiderata as much as possible. For the proposed $t$-prior, we recommend that researchers who would like to retain information consistency choose $\kappa \in (0, 1]$. For instance, setting $\kappa = 1$ results in a Cauchy prior. Note that, crucially, information consistency still holds if this Cauchy prior is centered on a value other than zero which enables one to incorporate expert knowledge about effect size by shifting the prior away from zero.
    Researchers who want to retain predictive matching should specify the prior to be centered on zero (i.e., $\mu_{\delta} =0$); however, the scale parameter $\gamma$ and the degrees of freedom $\kappa$ can be chosen freely. Next, we demonstrate with an example how the proposed Bayes factor can be used in practice. The example features a prior elicitation effort \citep[e.g.,][]{KadaneWolfson1998} highlighting the practical feasibility of specifying an informed prior based on expert knowledge.

	\section{PRACTICE}
	The \emph{facial feedback hypothesis} states that affective responses can be influenced by one's facial expression even when that facial expression is not the result of an emotional experience. In a seminal study, \citet{strack1988inhibiting} found that participants who held a pen between their teeth (inducing a facial expression similar to a smile) rated cartoons as more funny on a 10-point Likert scale ranging from 0-9 than participants who held a pen with their lips (inducing a facial expression similar to a pout).
	
	In a recently published Registered Replication Report \citep{wagenmakers2016registered}, 17 labs worldwide attempted to replicate this finding using a preregistered and independently vetted protocol. A classical random-effects meta-analysis yielded an estimate of the mean difference between the ``smile'' and ``pout'' condition equal to $0.03$ [95\% CI: $-0.11, 0.16$]. Furthermore, one-sided default Bayesian unpaired $t$-tests (using a zero-centered Cauchy prior with scale $1/\sqrt{2}$ for effect size, the current standard in the field of psychology; see \citealp{MoreyRouderBayesFactorPackage}) revealed that for all 17 studies, the Bayes factor indicated evidence in favor of the null hypothesis and for 13 out of the 17 studies, the Bayes factor in favor of the null was larger than 3. Overall, the authors concluded that ``the results were inconsistent with the original result'' \citep[p. 924]{wagenmakers2016registered}.
	
	Here we present an informed reanalysis of the data of one of the labs based on a prior elicitation effort with Dr.~Suzanne~Oosterwijk, a social psychologist at the University of Amsterdam with considerable expertise in this domain. The results for the other labs can be found in online appendix~C.
	
	\subsection{Prior elicitation}
	Before commencing the elicitation process, we asked our expert to ignore the knowledge about the failed replication of \citet{strack1988inhibiting}. Next, we stressed that the goal of the elicitation effort was to obtain an informed prior distribution for $\delta$ \emph{under the alternative hypothesis} $\mathcal{H}_1$, that is, under the assumption that the effect is present. This was important in order to prevent unwittingly eliciting a prior that is a mixture between a point mass at zero and the distribution of interest. Then, we proceeded in steps of increasing sophistication. First, together with the expert we decided that the theory specified a direction, implying a one-sided hypothesis test. Next, we asked the expert to provide a value for the median of the effect size: this yielded a value of $0.35$. Subsequently, we asked for values for the 33\% and 66\% percentile of the prior distribution for the effect size: this yielded values of $33\%\text{-tile} = 0.25$ and $66\%\text{-tile} = 0.45$. To finesse and validate the specified prior distribution we used the MATCH Uncertainty Elicitation Tool (\url{http://optics.eee.nottingham.ac.uk/match/uncertainty.php}; see also online appendix~B), a web application that allows one to elicit probability distributions from experts \citep{morris2014webbased}. Furthermore, we used \texttt{R}'s \citep{R} plotting capabilities for eliciting the prior number of degrees of freedom. The complete elicitation effort took approximately one hour and resulted in a $t$-distribution with location $0.350$, scale $0.102$, and 3 degrees of freedom. As shown in the theory part, this prior choice has a departure from the predictive matching criterion of $\pm 0.0198$ and misses information consistency by two samples. It should be emphasized, however, that the goal of this prior elicitation was to construct a prior that truly reflects the expert's knowledge without being constrained by considerations about Bayes factor desiderata. Alternatively, in an elicitation effort that puts more emphasis on these desiderata, one could, for instance, fix the degrees of freedom to one and let the expert only choose the location and scale.
	
	\subsection{Reanalysis of the Oosterwijk replication study}
	Having elicited an informed prior distribution for $\delta$ under the alternative hypothesis, we now turn to a detailed reanalysis of the facial feedback replication attempt from Dr.~Oosterwijk's lab at the University of Amsterdam. This data set features 53 participants in the ``smile'' condition with an average funniness rating of 4.63 ($SD = 1.48$), and  57 participants in the ``pout'' condition with an average funniness rating of 4.87 ($SD = 1.32$); consequently, the observed $t$ statistic is $t(108) = -0.90$.
	
	The alternative hypothesis is directional, that is, the teeth condition is predicted to result in relatively high funniness ratings, not relatively low funniness ratings. In order to respect the directional nature of the alternative hypothesis the two-sided informed $t$-test outlined above requires an adjustment. Specifically, the Bayes factor that compares an alternative hypothesis that only allows for positive effect size values to the null hypothesis can be computed via a simply identity that exploits the transitive nature of the Bayes factor \citep{morey2014simple}:
	\begin{equation}
		\label{informedEqDirectionalBF}
		\text{BF}_{+0}(d) = \underbrace{\frac{p(d \, | \, \mathcal{H}_{+})}{p(d \, | \, \mathcal{H}_{1})}}_{\text{BF}_{+1}(d)} \underbrace{\frac{p(d \, | \, \mathcal{H}_{1})}{p(d \, | \, \mathcal{H}_{0})}}_{\text{BF}_{10}(d)} = \text{BF}_{+1}(d) \text{BF}_{10}(d).
	\end{equation}
	We already showed how to obtain $\text{BF}_{10}(d)$, that is, the Bayes factor for the two-sided test of an informed alternative hypothesis; the correction term $\text{BF}_{+1}(d)$ can be obtained by simply dividing the posterior mass for $\delta$ larger than zero by the prior mass for $\delta$ larger than zero.\footnote{The expression for the marginal posterior distribution for $\delta$ is provided in Corollary~A.2.3 in the online appendix. Using this expression, numerical integration can be used to obtain the desired posterior mass.} The Bayes factor hypothesis test that we report will respect the directional nature of the facial feedback hypothesis and include the correction term from Eq.~\ref{informedEqDirectionalBF}.
	
	\begin{figure}
		\centering
		\includegraphics[width = 0.7 \textwidth]{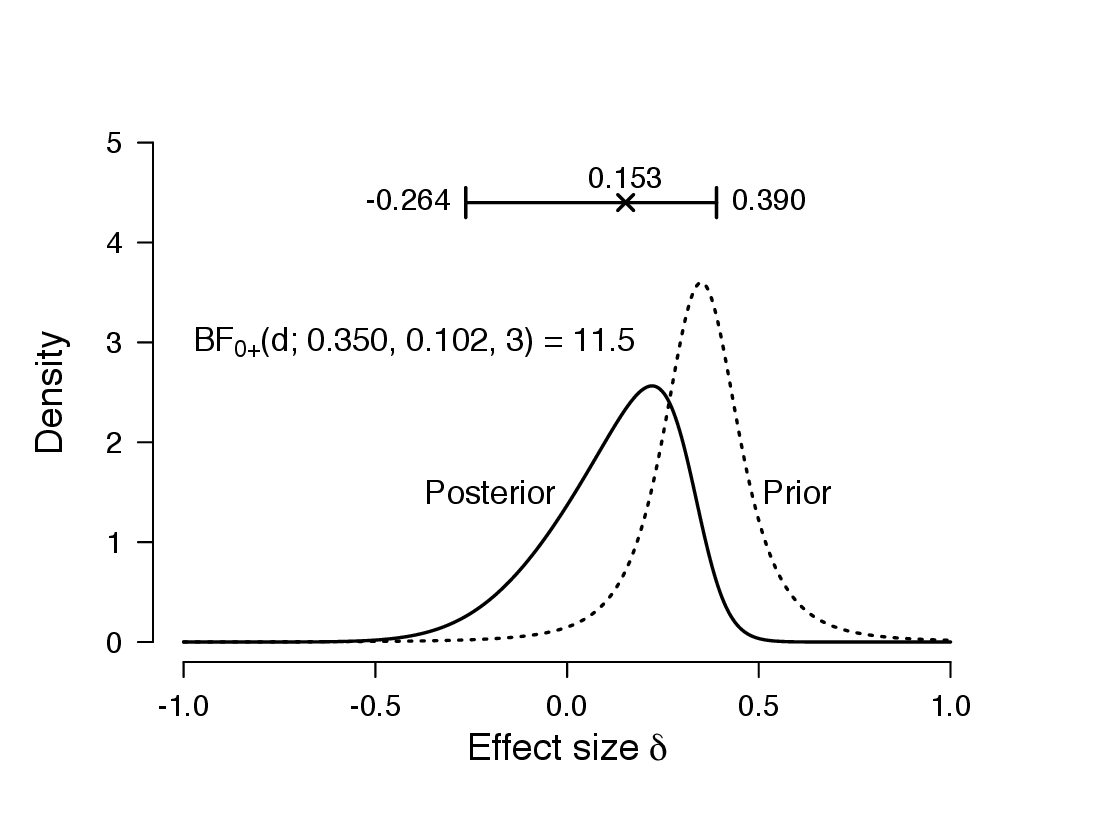}
		\caption{Results of an informed reanalysis of the facial feedback hypothesis replication data from the Oosterwijk lab. The dotted line corresponds to the elicited $\frac{1}{0.102}T_{3}\left( \tfrac{\delta - 0.350}{0.102}\right)$ prior distribution. The solid line corresponds to the associated posterior distribution, with a 95\% credible interval and the posterior median displayed on top. The Bayes factor in favor of the null hypothesis over the one-sided informed alternative hypothesis equals $\text{BF}_{0+}(d; 0.350, 0.102, 3) = 11.5$. Figure available at \protect \url{https://tinyurl.com/mk7uaxm} under CC license \protect \url{https://creativecommons.org/licenses/by/2.0/}.}
		\label{informedFigFacialPostInformed}
	\end{figure}
	
	Fig.~\ref{informedFigFacialPostInformed} shows the results of the reanalysis of the data from the Oosterwijk lab. The displayed prior and posterior distribution do not impose the directional constraint. The one-sided Bayes factor based on the informed prior equals $\text{BF}_{0+}(d; 0.350, 0.102, 3) = 11.5$, indicating that the data are about twelve times more likely under the null hypothesis than under the one-sided alternative hypothesis.
	
	\begin{figure}
		\centering
		\includegraphics[width = 0.7 \textwidth]{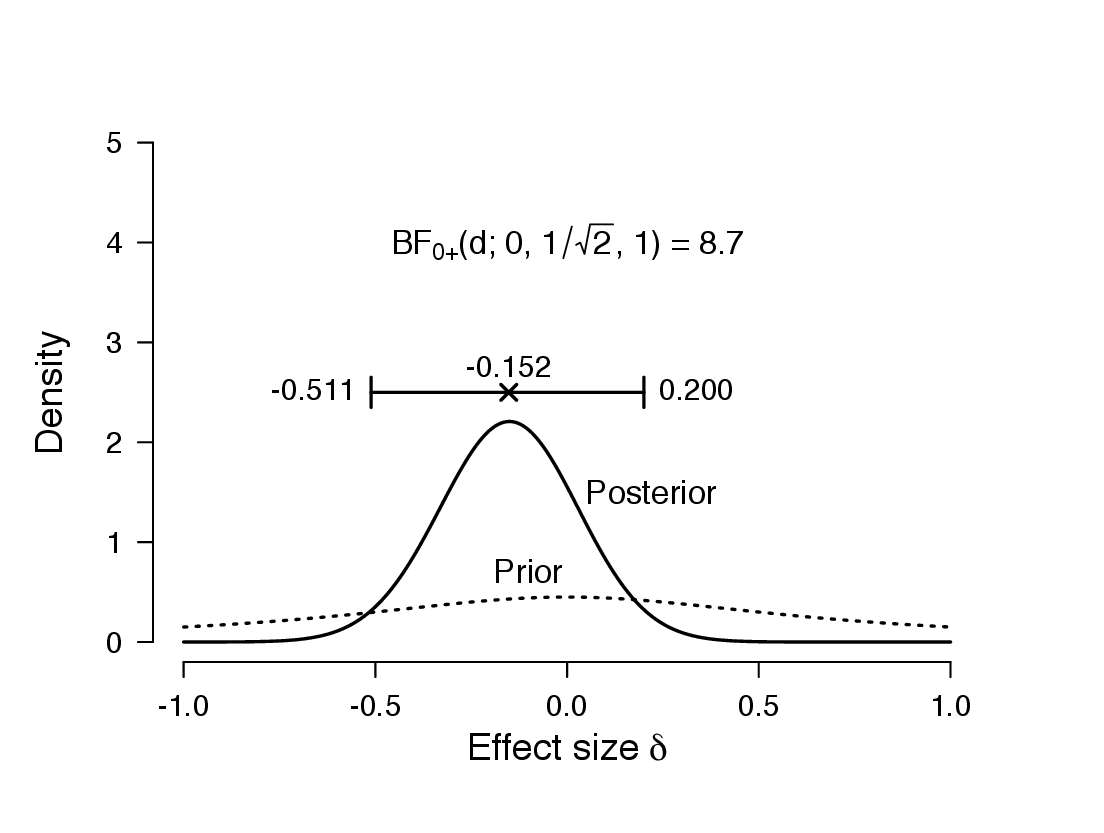}
		\caption{Results of the default analysis of the facial feedback hypothesis replication data from the Oosterwijk lab. The dotted line corresponds to the default Cauchy prior distribution with scale parameter $1/\sqrt{2}$. The solid line corresponds to the associated posterior distribution, with a 95\% credible interval and the posterior median displayed on top. The Bayes factor in favor of the null hypothesis over the one-sided default alternative hypothesis equals $\text{BF}_{0+}(d; 0, 1/\sqrt{2}, 1) = 8.7$. Figure available at \protect \url{https://tinyurl.com/mgs28ob} under CC license \protect \url{https://creativecommons.org/licenses/by/2.0/}.}
		\label{informedFigFacialPostDefault}
	\end{figure}
	
	For comparison, Fig.~\ref{informedFigFacialPostDefault} displays the results based on the default one-sided zero-centered Cauchy distribution with scale $1/\sqrt{2}$. The one-sided default Bayes factor equals \sloppy$\text{BF}_{0+}(d; 0, 1/\sqrt{2}, 1) = 8.7$, indicating that the data are about 9 times more likely under the null hypothesis than under the one-sided default alternative hypothesis. Hence, both the informed and the default Bayes factor yield the same qualitative conclusion, that is, evidence for the null hypothesis. However, the unrestricted posterior distributions differ noticeably between the informed and the default analysis: the posterior median based on the informed prior specification is positive and equal to $0.153$ (95\% credible interval: $[-0.264, 0.390]$) whereas the posterior median based on the default prior distribution is equal to $-0.152$ (95\% credible interval: $[-0.511, 0.200]$).
	
	\section{CONCLUDING COMMENTS}
	The comparison between two means is a quintessential inference problem.
	Originally developed by \citet{jeffreys1948theory2} in the one-sample setting, the Bayesian $t$-test has recently been extended to the two-sample set-up by \citet{gonen2005bayesian} and, subsequently, by \citet{rouder2009bayesian} and \citet{wang2016simple}.
	Here we showed that practitioners can easily and intuitively use a generalized version of the Bayes factor by \citet{rouder2009bayesian} to inform their two-sample Bayesian $t$-tests. We used the framework of \citet{gonen2005bayesian} and extended the priors by \citet{rouder2009bayesian} to allow for more informed Bayesian $t$-tests that can incorporate expert knowledge by using a flexible  $t$-prior.  
	An advantage of the flexible $t$-prior is that it contains the objective default prior by \citet{rouder2009bayesian} as a special case and the subjective prior proposed by \citet{gonen2005bayesian} as a limiting case.
	Therefore, practitioners can use the same formula to compute subjective and objective Bayesian $t$-tests.	
	To encourage its adoption in applied work, we have implemented the proposed Bayesian $t$-test set-up in the open-source statistical program \texttt{JASP} \citep[\url{jasp-stats.org}]{JASP2018}.
	In the theoretical part of this article, we investigated theoretical properties of the informed $t$-prior. Specifically, we discussed popular Bayes factor desiderata and proposed measures to quantify the deviation of an informed $t$-test from its objective counterpart.
	In the practical part of the article, we illustrated the use of the informed Bayes factor with an example.
	Similar to the prior proposed by \citet{gonen2005bayesian}, the flexible $t$-prior may encourage the use of prior distributions that better represent the predictions from the hypothesis under test, allowing more meaningful conclusions to be drawn from the same data (\citealp{rouder2016is}, \citeyear{rouder2016interplay}).
		
	Other choices than a $t$-prior for effect size are conceivable. Eq.~\ref{eqGonen} shows that one can obtain a Bayes factor for any scale-mixture of normals by integrating Eq.~\ref{eqGonen} with respect to a prior on $g$ (see Theorem~A.3 in the online appendix; for possible choices see, e.g., \citealp{liang2008mixtures} and \citealp{bayarri2012criteria}). This also includes the prior proposed by \citet{wang2016simple} and highlights that it is straightforward to extend this prior to include a location parameter that can be specified based on expert knowledge.
	In fact, the expressions for the Bayes factor that we presented make it relatively straightforward to use \emph{any} proper prior on standardized effect size (see Eq.~\ref{eq:BF10}). The proposed departure measures can then be used to investigate information consistency and predictive matching for different choices.
	
	In this article, we focused on the Bayes factor as the inferential tool for quantifying the relative evidence for competing hypotheses based on observed data. However, it could be argued that a complete Bayesian analysis requires one to also specify the prior plausibilities of the competing hypotheses. This is of particular importance in situations where unlikely hypotheses are tested or when multiple comparisons are considered \citep{ScottBerger2010}. Although specifying the prior plausibilities of the competing hypotheses may not be trivial, once this has been achieved, the Bayes factor can be simply multiplied by the prior odds to obtain the posterior odds of interest.
	
	\bigskip
	\begin{center}
		{\large\bf SUPPLEMENTARY MATERIAL}
	\end{center}
	
	\begin{description}
		\item Online Appendix: Informed Bayesian $T$-Tests: Derivations, details about prior elicitation, and additional analyses. (pdf)
	\end{description}
	
	\bibliographystyle{apalike}
	\bibliography{Alexander,references}
\end{document}